# Experimental study of the interaction of two laser-driven radiative shocks at the PALS laser

R. L. Singh[a,d,*], C. Stehlé[a], F. Suzuki-Vidal[b], M. Kozlova[c,f], J. Larour[d], U. Chaulagain[a,e], T. Clayson[b], R. Rodriguez[j], J.M. Gil[j], J. Nejdl[c,f], M. Krus[f,c], J. Dostal[f,c], R. Dudzak[c,f], P. Barroso[g], O. Acef[h], M. Cotelo[i], P. Velarde[i]

[a]*LERMA, Observatoire de Paris, UPMC, CNRS, ENS, UCG, Paris, France*
[b]*Imperial College London, London, UK*
[c]*Institute of Physics of the Czech Academy of Sciences, Na Slovance 1999/2, 182 21Prague, Czech Republic*
[d]*LPP, CNRS, Ecole polytechnique, UPMC Univ Paris 06, Univ. Paris-Sud, Observatoire de Paris, Universit Paris-Saclay, Sorbonne Universits, PSL Research University, 75252Paris, France*
[e]*ELI Beamlines, Institute of Physics ASCR, Na Slovance 1999/2, Prague, 182 21, CzechRepublic*
[f]*Institute of Plasma Physics of the Czech Academy of Sciences, Za Slovankou 1782/3, 18200 Prague, Czech Republi*
[cg]*GEPI, Observatoire de Paris, PSL Research University, CNRS, Université Paris Diderot, Sorbonne Paris Cité, Paris, France*
[h]*SYRTE, Observatoire de Paris, UPMC, CNRS, Paris, France*
[i]*Instituto de Fusion Nuclear, UPM, Madrid, Spain*
[j]*Universidad de las Palmas de Gran Canaria, Las Palmas, Spain*

## Abstract

Radiative shocks (RS) are complex phenomena which are ubiquitous in astrophysical environments. The study of such hypersonic shocks in the laboratory, under controlled conditions, is of primary interest to understand the physics at play and also to check the ability of numerical simulations to reproduce the experimental results.

In this context, we conducted, at the Prague Asterix Laser System facility (PALS), the first experiments dedicated to the study of two counter-propagating radiative shocks propagating at non-equal speeds up to 25-50 km/s in noble gases at pressures ranging between 0.1 and 0.6 bar. These experiments highlighted the interaction between the two radiative precursors. This interaction is qualitatively but not quantitatively described by 1D simulations. Preliminary results obtained with XUV spectroscopy leading to the estimation of shock temperatureand ion charge of the plasma are also presented.

*Email address:* raaj.phys@gmail.com *(R. L. Singh)*

## 1. Introduction

Radiative shocks are strong shocks (i.e. the Mach number, M>>1), which reach high temperatures and thus are the source of intense radiation [1, 2,3]. Depending on the opacity, the radiation emitted from the shock may be absorbed by the pre-shock region, inducing its pre-heating. Such pre-heated zone is termed as the radiative precursor (i.e. a radiative ionization wave) [1, 4, 5]. Radiative shock waves have been studied experimentally since more than a decade, mostly on large-scale laser facilities [4, 5, 6, 7, 8, 9, 10, 11, 12, 13], in noble gases and with different targets geometries. With laser intensities on the target comprised between $10^{14}$ and $10^{15}$ W/cm$^2$, these experiments allowed to record shock speeds ranging between 40 and 150 km/s.

Tubular targets have been used in many shock experiments. Hence, in this case, the shock tends to fill the tube and in a first approximation, may be assumed to behave as 1D. The experiments can be then used for code benchmarking. In this configuration, several studies have been focused on the characterization of the radiative precursor [4, 8, 6] for shock speeds ∼ 60 km/s. In such experiments, electron densities up to $10^{19}$ cm$^{-3}$ have been recorded by visible interferometry. The radiative losses at the tube boundaries have been pointed out leading to a strong diminution of the electron density when compared to the results from 1D simulations [4, 14, 6]. Such radiative losses depend on the walls material and have been estimated to be 40% for Aluminum and Silica [6]. The losses lead to a small curvature of the ionization front and to a reduction of its longitudinal extension [14, 15]. At higher speeds (∼ 200 km/s), x-ray radiography pointed out a collapse of the post-shock [7] due to the radiation losses. Finally, for these high-speed conditions, the wall heating leads to the development of secondary wall shocks, which interact with the primary shock [16], and which have not been observed at lower speeds.

Contrary to the case discussed above, if the shock wave does not fill the tube, 2D effects are more pronounced as shown in a recent experiment dedicated to XUV imaging of both the post shock and the radiative precursor of a RS wave propagating at 45 km/s in Xenon at 0.3 bar [5].

All previous experimental studies have been focused on the case of isolated radiative shocks. However, in astrophysical conditions, the radiative shock often interacts with a denser medium, leading to the development of reflected and transmitted shocks. A few representative examples of such phenomena are the interaction of supernovae remnants with dense molecular clouds [17, 18], the accretion shocks on the photosphere of T-Tauri stars [19] and the bow shocks at the head of stellar jets [20, 21]. The collision (or the interaction) of two radiative shock waves is obviously a rare astrophysical event and the template case of supernova remnant DEM L316 (see fig. 1 of [22]) is still the subject of debates [23, 24, 25] as the observation of these two different shocks can be also interpreted as the superposition of two blast waves in the field of view of the telescope. In this context, the development of dedicated laboratory experimentsto the study of propagation and interaction of counter-propagating shock waves is important to characterize such events through their specific signatures.

In this paper, we present the results of experiments performed at the Prague Asterix Laser System (PALS) facility [26] on the study of the interaction of two radiative shock waves in Xenon at low pressure ($< 1$ bar). These shock waves are launched by two laser beams with different energy and wavelength and therefore the shock waves have different speeds, which are comprised between $\sim 20$ and $50$ km/s.

Section 2 presents numerical studies of the interaction of two shock waves with identical (50 - 50 km/s) and different (50 - 20 km/s) speeds. The experimental setup is then presented in section 3. It includes a description of the twomain diagnostics namely, time-dependent optical laser interferometry, to probe the radiative precursors before the collision time, and time and space integrated XUV spectroscopy, to derive estimates of the electron temperature through relevant spectral signatures. Section 4 discusses the results derived by these diagnostics. Concluding remarks are presented in the last section.

2. Interacting shock waves

We investigate here the characteristic parameters and dynamics of two counter-streaming shocks, and of a single shock, through 1D simulations. These simulations were performed employing the Lagrangian numerical code 'HELIOS', using the associated PROPACEOS equation of state and opacity [27]. For the opacity, we have used, for the Xe gas, a multiplier $\times 20$ (to adjust with our own opacities [28], see Appendix). For our qualitative study, the number of groups is set to be 1. The target cell, with a 4-mm length, is filled with Xe at 0.1 bar. Two gold coated-CH foils (total thickness 11.6 µm) are placed at both ends closing the cell.

We present the results of three representative sets of simulations performed in Xe at 0.1 (mass density = $5.4 \times 10^{-4}$ g/cm$^{-3}$) bar namely, (I): for a single RS at ∼ 50 km/s moving from the left end of the target cell to the right end, as well as for two identical RS at ∼ 50 km/s propagating in opposite directions (i.e. starting from the left and right end, respectively, Fig.1), (II): for the same conditions but without any coupling with radiation (Fig.2) and, (III): for two counter-propagating radiative shocks of different speeds in which, one shock is propagating with the speed of ∼ 50 km/s from the left end of the cell while another shock propagates with the speed of ∼ 20 km/s from the right end of the cell (Fig. 3). To achieve the aforementioned speeds in the simulation, we have used on the left and right sides two laser beams ($\lambda$ = 438 nm) with the adequate fluences. The pulse duration is set to 0.3 ns (peak at 0.15 ns), to reproduce the experimental conditions detailed later in the paper.

Firstly, we discuss the case of two counter-propagating identical RS. Fig.1 shows the variations of the electron density ($N_e$) and temperature ($T_e$) in the Xenon layers. This academic case is fully symmetrical and it is equivalent to the case of one RS with a fully reflective boundary (for hydrodynamics and radiation) in the middle of the tube. The two shocks appear in Xenon at ∼ 2 ns and the collision occurs at ∼ 38 ns. At 3 ns, the precursor extension is ∼ 0.08 cm, whereas the post-shock electron density and electron temperature are $7.8 \times 10^{20}$ cm$^{-3}$ and 14 eV respectively. The length of precursor increases rapidly with time and the two precursors merge suddenly at ∼ 8 ns. As the time progresses, the merging effect increases significantly. It is characterized by a flat common precursor, whose electron density and electron temperature are increasing with time. At the time of the collision (∼ 38 ns), the post-shock mass density and electron density jump from 0.011 to 0.14 g/cm$^{-3}$ and $6.7 \times 10^{20}$ to $6.6 \times 10^{21}$ cm$^{-3}$, whereas the electron temperature rises up to 39 eV. The collision leads to the development of two reverse shock waves propagating back in Xenon and in the different layers of the piston, with a speed of 15 km/s. These reverse shocks lead to a dense plasma ($N_e > 10^{21}$ cm$^{-3}$) which is not accessible to the present experiment and will not be detailed here.

To investigate the effects of the interaction, we compare the previous results with those obtained for a single radiative shock, moving from the left to the right direction in the cell (plotted by dotted lines in Fig 1). The single shock propagates identically to the case (I) until 10 ns. After this time, the profiles of the electron temperature and density differ strongly from the previous case and their values are lower than for the 2RS. The post-shock extension is slightly smaller than for the 2RS. This last effect is due to the fact that the shock wave propagates in a warmer medium, which modifies the opacity and the sound speed. In order to highlight the effect of the radiation, we have performed another simulation with the same set of parameters as of above, however, putting the Xenon opacity equal to zero. The result of the simulation is presented in Fig.2. The collision time is now 40 ns and the post-shock is no more compressed by radiation cooling. Its compression at 10 ns is 10 instead of 35. There is no radiative precursor. Moreover, there are no differences in the $N_e$, and $T_e$ profiles of the single shock and that of the two counter-propagating shocks before the collision time.

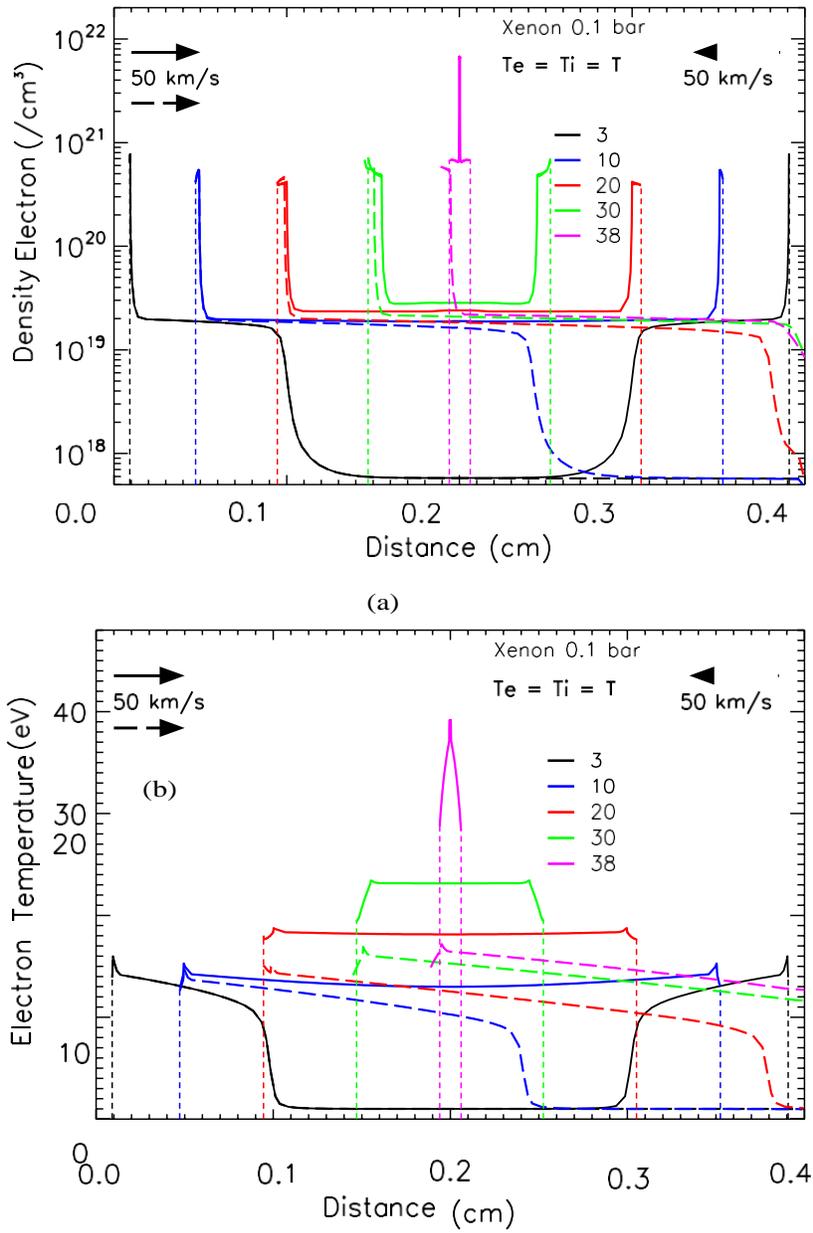

Figure 1: Electron density N$_e$ (a) and electron temperature T$_e$ (b) versus axial position (along a 0.4 cm long shock tube) at 3, 10, 20, 30 and 38 ns from HELIOS simulations for the cases of single shock of ∼ 50 km/s (dashed line) and two identical counter-propagating shocks of ∼ 50 km/s (solid lines). The vertical dotted lines show the position of the interface between piston and backing Xenon gas.

In the set (III), we have performed the simulations for the case of two counter-propagating shocks of different speeds, i.e. 50 and 20 km/s starting from the left and the right ends of the cell, respectively. Like previously, we have also performed the simulation for the two single shocks of respective speeds 50 (from the left end) and 20 km/s (from the right end). This more complicated case study is primarily motivated by our experimental setup. However, it is interesting to compare these results with the simpler previous case of two counter-streaming identical shocks. The spatial and temporal variations of $N_e$ and $T_e$, are plotted at times 3, 10, 30, 38 and 49 ns in Fig. 3a and 3b.

The left and the right shocks appear in Xenon at ∼ 2 and 3 ns, respectively. Later, at 10 ns, the two precursor extensions are respectively equal to 0.18 and 0.034 cm. The merging of the two precursors starts at ∼ 15 ns. As expected from the values of shock speeds, the collision time occurs at 49 ns, which is delayed in comparison to the case (I). It may be noted that, up to this collision time, the post-shock conditions are identical for the cases of a single RS at 50 km/s and the present left shock at the same speed. This reveals that there is no noticeable effect of the right RS with speed 20 km/s on the left RS of 50 km/s. On the contrary, we note a difference in extension of the right post shock from the counter-propagating case compared with the single shock propagating from the right at 20 km/s (Fig. 3a). The two radiative precursors merging results in a plateau for the electron density and temperature. The temperature at collision time is now 28 eV instead of 39 eV in the case I, and the electron density reaches up to $3.1 \times 10^{21}$ cm$^{-3}$ instead of $6.6 \times 10^{21}$ cm$^{-3}$. This numerical study gives the main characteristics of the interaction of two counter-propagating shock waves propagating in Xenon at 0.1 bar with speeds equal to 50-50 km/s and 50-20 km/s. The case of identical speeds is simpler due to the symmetry of the problem. However, whatever the speeds, the most important signature of the interaction is the merging of the precursor at 8 ns for 50-50 km/s and at 15 ns for 50-20 km/s. This merging is followed by a regular increase with time of the electron density and the temperature. The collision time is characterized by a sudden increase of the electron density by an order of magnitude, reaching 6.6 × $10^{21}$ and 3.1 × $10^{21}$ cm$^{-3}$ respectively, whereas the temperature increases up to 39 and 28 eV.

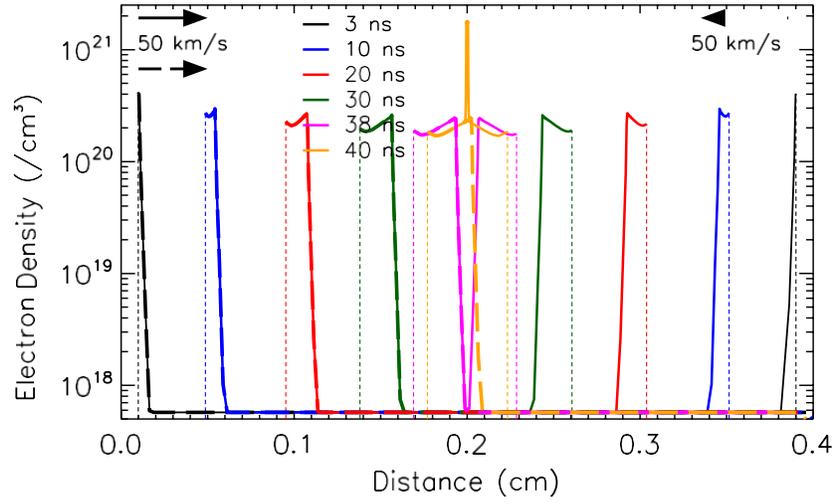

(a)

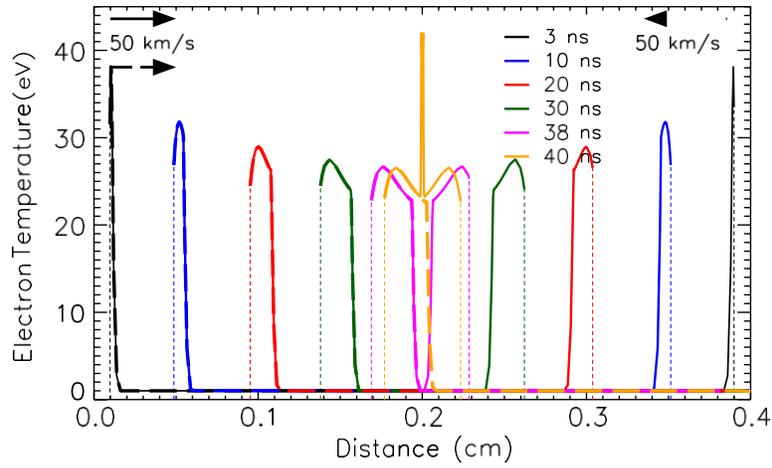

(b)

Figure 2: Variations of $N_e$ (a) and $T_e$ (b) versus axial position for the case of two identical counter-propagating shocks (of speeds ∼ 50 km/s) at 3, 10, 20, 30, 38 and 40 ns as derived from HELIOS simulations. For these simulations we have neglected the effect of radiation in HELIOS by keeping the opacity to be zero. The vertical dotted lines show the position of the interface between the piston and backing Xenon gas.

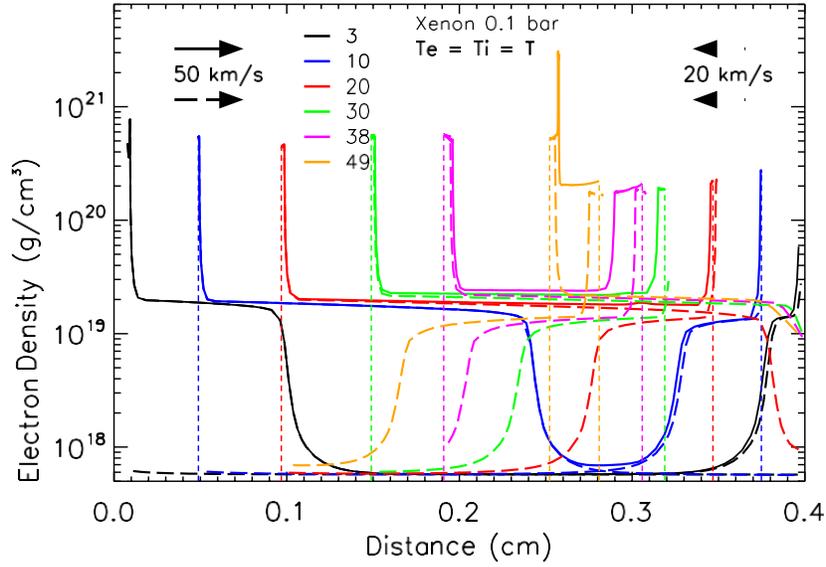

(a)

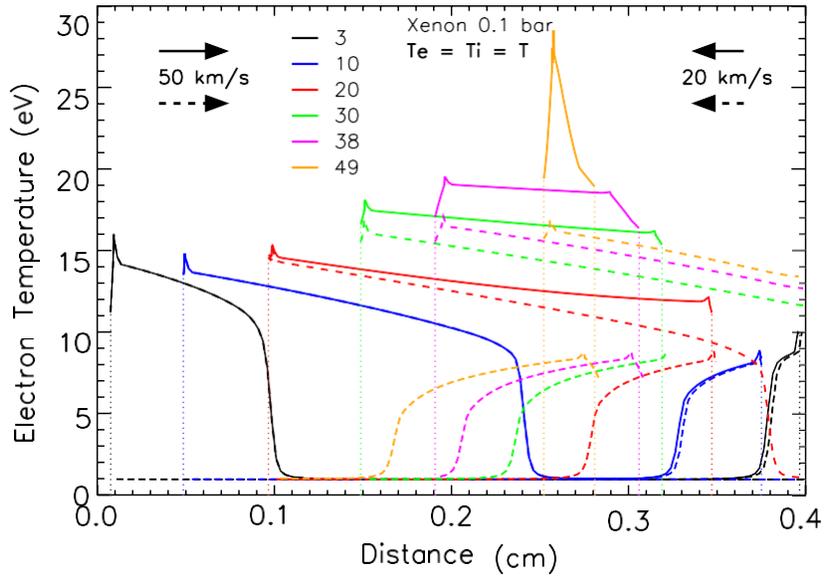

(b)

Figure 3: Variations of N$_e$ (a) and T$_e$ (b) with axial position for the case of two non-identical counter-propagating shocks (of speeds ∼ 50 & 20 km/s) and two single shock (dotted lines)of speeds ∼ 50 & 20 km/s respectively at 3, 10, 30, 38 and 49 ns as derived from HELIOS simulations. The vertical dotted lines show the position of the interface between piston and backing Xenon gas.

## 3. Experimental Setup

As mentioned in the previous section, the two laser beams, which have been used to drive the two counter streaming shocks are not identical and thus will drive shock waves at different speeds. The first beam, at 438 nm, also termed as MAIN laser beam, has a nominal energy of ∼ 120 J (measured on target) whereas the second laser beam, at 1315 nm, hereafter AUX laser beam, has a lower energy ∼ 60 J (measured before the entrance window of the vacuum chamber). The two laser beams are focused by lenses (f = 564 mm for MAIN and 1022 mm for AUX) on the opposite sides of the millimetric sized target in the vacuum chamber.

Two phase plates are placed before each laser lens. The MAIN phase plate, successfully used in previous experiments [6, 8] is designed to produce a uniform intensity distribution over a circular section of diameter 0.5 mm, whereas it is 0.25 mm for the AUX phase plate. The expected fraction of the laser energy in the focal spots is about 80%. Two keV pinhole cameras are employed to monitor the laser's impact on the two pistons. These time integrated records provide estimates of the laser spot diameters ranging between 500 - 600 $\mu$m for MAIN, and 250 - 300 $\mu$m for AUX.

*3.1. Targets*

The targets are placed inside the PALS vacuum chamber and filled in situ with Xe or Xe+He (90 -10 %) mixture at 0.1 - 0.6 bar. The targets consists of a channel of a parallel pipe shape having the dimension of $0.9 \times 0.6 \times 4$ mm, placed at the top of an aluminum structure (Fig. 4). The channel is closed laterally by two fused silica ($SiO_2$) windows of 500 $\mu$m thickness. From the top, it is closed with a 100-nm thin $Si_3N_4$ membrane, which is supported by a silicon frame and is transparent to XUV radiation (Fig. 4). Two foils, made of Parylene-N (11 $\mu$m) and coated with Au (0.6 $\mu$m) are placed at both ends to close the channel (Fig. 5) to act as pistons to drive the shocks from both sides inside the target. Whereas the Parylene layer will be ablated by the laser, the gold coating aims at blocking the x-rays generated during ablation, preventing the preheating of gas in the channel. As a consequence of the ablation process, the two foils will be propelled in the tube. These foils are glued on 0.1 mm thick Nickel disks, which internal diameter of 1 mm. This disk aims at helping the assembling of the targets. It also contributes to prevent radiation from the laser impact on the foil to reach the gas in the tube.

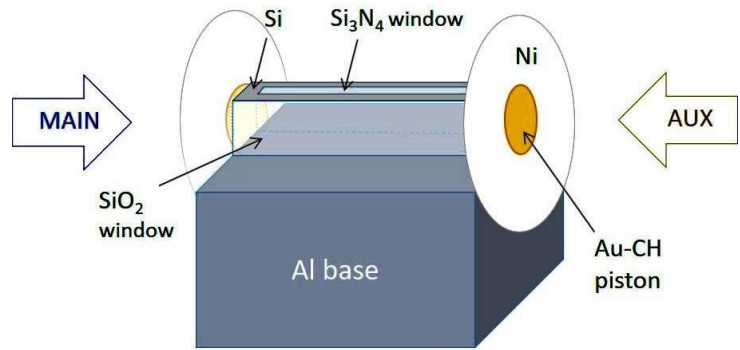

Figure 4: Schematics of the target showing the two Au-CH pistons closing the tube (on their respective Ni disks), the horizontal $Si_3N_4$ window its a Silicon frame and the two vertical $SiO_2$ windows which are supported by an Al base. Figure is not drawn to scale.

*3.2. Visible Interferometry*

A visible interferometry diagnostic is used to estimate the speed of the shocks and the precursor electron density. The Mach-Zehnder setup is shown at the bottom of the chamber in Fig. 5. The target is placed in one arm of the interferometer (Fig. 5) and is illuminated by a probing EVOLUTION laser (beam diameter $\sim$ 1 cm, wavelength 527 nm, duration $\sim$ 200 ns). The fringes are recorded on a HAMAMATSU C7700 visible streak camera (CCD pixel size 24 $\mu$m). The target is imaged on the slit of this camera with a magnification of 2.26. We thus record 2D images having the position along the slit versus time. The sweeping is kept either 50 or 200 ns, and the slit opening is set to 200 $\mu$m (100 $\mu$m on the target). In the *longitudinal* configuration, the horizontal line which connects the centers of the MAIN and AUX laser spots is imaged on the slit of the streak camera (i.e. nominally at a distance of 300 $\mu$m from the base of the channel). In the *transverse* mode, the slit records the images of a vertical line located at a given distance from the pistons. This is done using a Dove prism, which allows making a rotation of the image by 90 degrees.

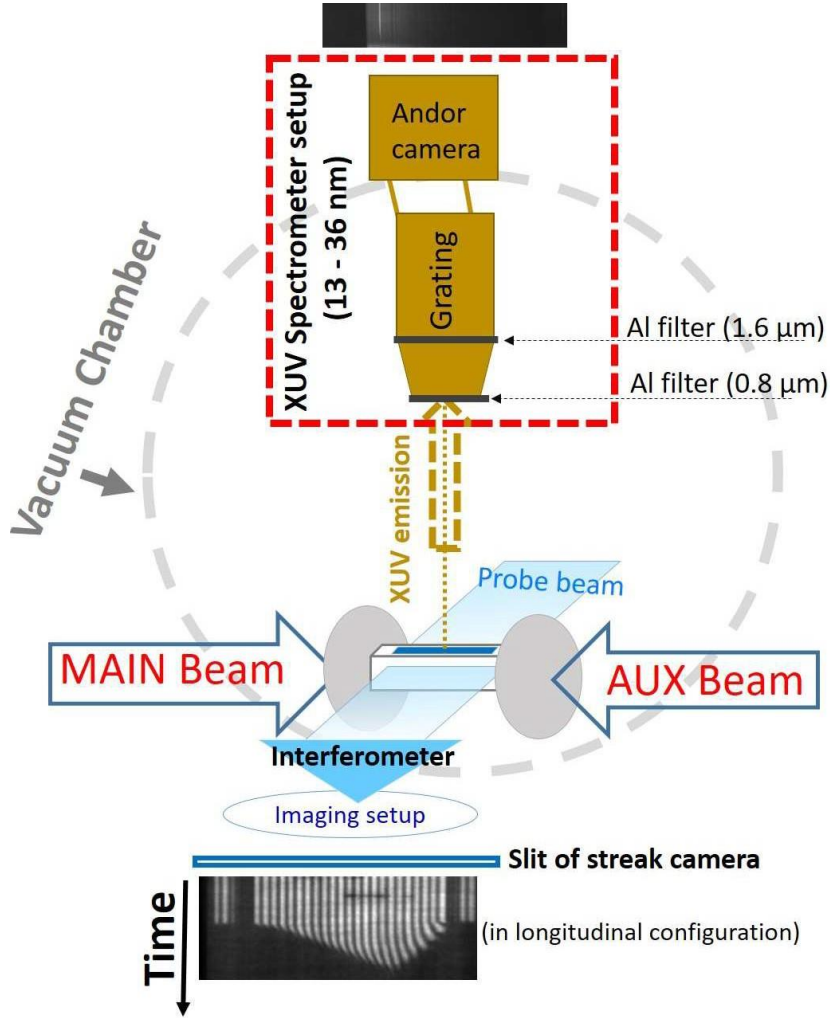

Figure 5: Schematic of experimental setup. MAIN beam: E ~ 120 J at 438 nm and AUXbeam: E ~ 60 J at 1315 nm.

*3.3. XUV Spectroscopy*

Time and space integrated XUV emission spectra between 15 and 35 nm have been recorded employing a flat field XUV spectrometer (shown inside the red dotted box in Fig. 5), using a cylindrical gold coated grating (curvature radius of 5649 mm, 1200 groves per mm, efficient grating area of 45 × 27 mm, blaze angle 3.7 degrees). This spectrometer is installed on the top of the vacuum chamber, facing the $Si_3N_4$ window. A cooled Andor DX440 CCD is attached tothe spectrometer to record the spectrum of the XUV radiation. Two Al filters of thickness 0.8 and 1.6 $\mu$m, respectively protect the grating and the CCD camera from debris and visible light. The wavelength calibration is performed using the L - edge of Al filter at 17 nm in the first and second orders. The goal of the XUV spectroscopic analysis is to provide information about the plasma temperature. To make this interpretation easier, traces of Helium (10% in number) can be introduced in Xenon for some shots.

4. Results

The records obtained from the experiments have been processed to estimate the shock section, speed, electron temperature and density. We shall briefly present the results obtained with the transverse interferometry, as they provide qualitative information about the curvature and transverse extension of the radiative precursor. Then, we shall focus on the results of the longitudinal interferometry which enabled to measure the shock speed and the precursor electron density. The spectroscopic records will then complete this analysis with estimates of the electron temperature.

*4.1. Transverse Interferometry*

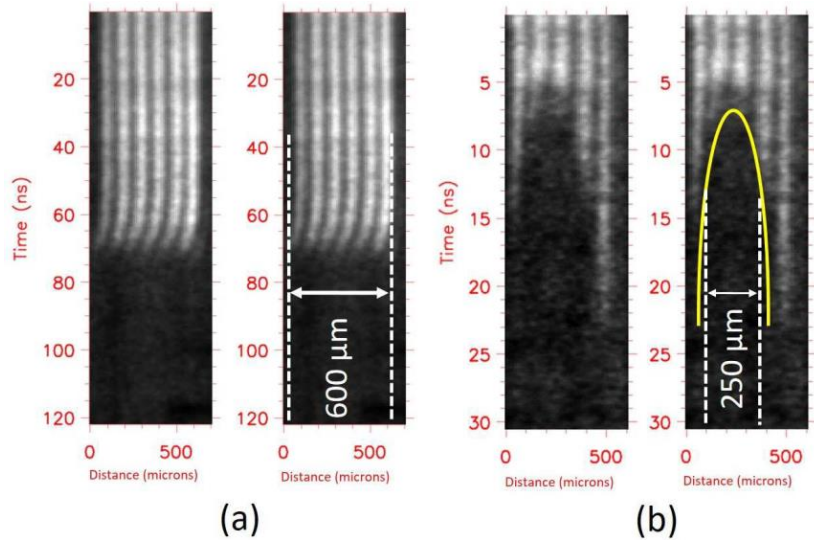

Figure 6: Transverse interferometric images for (a): shot#48111 (MAIN shock only). (b): shot#48130 (AUX shock only). The time is measured after an offset equal to 14 and 23 ns respectively. The position 'zero' on the x-axis of each image corresponds to the base of the target.

A transverse interferometric record for the MAIN shock alone in Xe at 0.2

bar is reported in Fig. 6(a). The setup here images on the camera a transverse section of the tube which is located at a distance d$_{slit}$ equal to 3 mm from the initial position of the MAIN piston. Taking into account the offset of 14 ns, the time of shock arrival is recorded at 72 ns after the time $t_0$ of laser arrival on the target and the shock speed is estimated to be ∼ 35 km/s. Due to the presence of glue on one lateral window (on the right part of the figure), only 6 fringes are visible. The lateral extension of the shock structure at this time is derived to be ∼ 600 μm in qualitative agreement with the specifications of the MAIN phase plate, and the structure of precursor is relatively flat. The axis of symmetry of the shock is estimated to be at ∼ 350 microns from the base of the target (i.e. 50 μm above the nominal value of 300 μm).

A record for the AUX shock alone is shown in Fig. 6(b), which corresponds to a gate opening of 50 ns. The start time of the image has an offset of +23 ns from $t_0$ and the distance d$_{slit}$ is set to 700 microns from the initial position of the AUX piston. The AUX shock duration extends from 30 ns to at least 34 ns after $t_0$. The shock speed is then estimated to be ranging between 23 and 20 km/s. The structure of the precursor is strongly bent. It may further be noted that the lateral spread of the shock is ranging between 250 - 300 μm (which is also in agreement with the specifications of the AUX phase plate) and that the axis of symmetry of the shock system is also located at about 350 μm from the bottom of the cell.

*4.2. Longitudinal Interferometry*

The interferometric images have been processed with a frequency filtering scheme in the Interactive Data Language (IDL) to enhance the fringes contrast. The locations of the maximum intensity in each fringe are used to derive the shock speed (Fig. 7) and the average electron density (Fig. 8) as presented below.

*4.2.1. Shock Speed*

A typical longitudinal interferometric record is shown in Fig. 7 for Xenon at 0.1 bar and two counter streaming shocks driven by laser beams with energies 133 and 68 J for the MAIN and the AUX shocks respectively. To derive the shocks speeds, we determine the position of the last visible end points of the fringes, where the electron density reaches the maximum value accessible to the diagnostic. The fringes are strongly bent, as expected due to the presence of the radiative precursors. We assume the shock front to be close to the location of these last visible end points. A linear regression of the connecting points is then used to measure the shock speed, which, in this case, is $54 \pm 3$ and $23 \pm 3$ km/s for MAIN and AUX shock waves, respectively. This speed determination is based on the absorption behavior of the plasma and not on the real position of the front discontinuity, which is not visible due to the strong absorption. The collision of the two shocks is located at $\sim 2800$ $\mu$m (t = 47 ns after the start of the shock propagation).

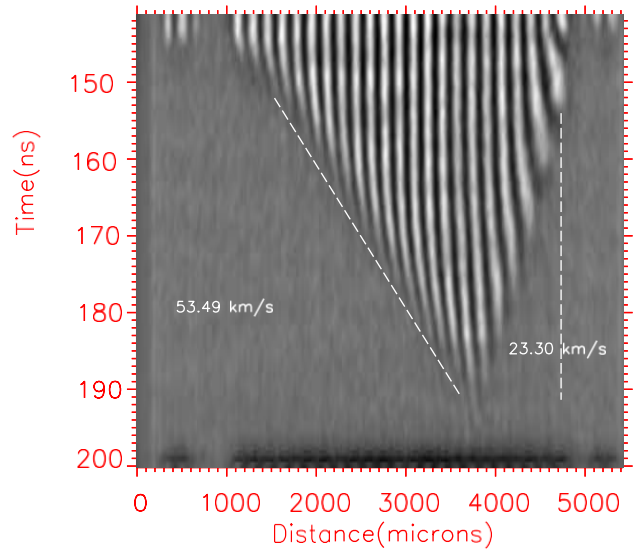

Figure 7: Interferometric image recorded for the shot #48055 in Xe at 0.1 bar. The shock speeds for the shocks driven by MAIN land AUX lasers are respectively equal to $\sim 54$ and 23 km/s. The time of laser arrival on the piston is at 146 ns. The positions of the Au-Xe interface on the record are respectively 954 and 4954 microns.

*4.2.2. Averaged electron density*

The average of the electron density along the path of the probe laser beam (*line averaged electron density*) in the plasma can be estimated from the record.

For this purpose, we compare the relative variation of the phase shift with time at each fringe maxima with its value at the time of laser arrival. The phase shifts allow to calculate the electron density using the following equation:

$$\Delta\emptyset = \frac{\pi d}{\lambda N_c} <N_e> \qquad \ldots\ldots (1)$$

where $\lambda = 527$ nm is the wavelength of the probing laser, $N_c = 4 \times 10^{21}$ cm$^{-3}$ the critical density at this wavelength, and $<N_e>$, is the electron densityaveraged over the laser path $d$ in the cell, defined as:

$$<N_e> = \int_0^d \frac{N_e(z,t)dy}{d} \qquad \ldots\ldots.. (2)$$

In the following, we shall take d = 600 $\mu$m, which corresponds to the horizontal transverse section of the shock channel. For a plasma which is uniform within this section, $<N_e>$ is close to the real local value of the electron density. As seen from the transverse interferometric record, this is rather true for the MAIN shock wave, whereas it is not the case for the AUX shock, which has a smaller section ($\sim$ 300 $\mu$m, see 4.1) and which is obviously not uniform.

| Shot number | Gas | Pressure (bar) | MAIN laser Energy (J) | AUX Shock Energy (J) | MAIN Shock Speed (km/s) | AUX Shock Speed (km/s) |
|---|---|---|---|---|---|---|
| 48055 | Xe | 0.1 | 133 | 68 | 54±3 | 23±3 |
| 48132 | Xe+He | 0.2 | 118 | 57 | 41±4 | 18±2 |
| 48138 | Xe+He | 0.2 | 121 | 0 | 45±4 | No shock |

Table 1: Experimental details of the interferometric records. The speed is deduced by the method of the last fringe explained in the text (see subsection 4.2.1).

To analyse the interaction of the shocks, we shall discuss below three representative interferometric records for Xenon. The conditions of these records are presented in Table 1 and the records are shown in Fig. 8. Two shots concern the propagation of two radiative shock waves at 0.1 and 0.2 bar. For comparison, one record is dedicated to the propagation of one shock (MAIN) only at 0.2 bar. The two records at 0.2 bar are performed in the Xe-He mixture (90 - 10% in numbers of atoms). At the precision of our records, we consider that this impurity concentration has only a negligible effect on the shock speed and the precursor electron density. The limit of detection of $<N_e>$ over the section of the tube (0.6 mm) is about $7 \times 10^{17}$ cm$^{-3}$ (corresponding to 2 pixels) for all the records.

Five colors (white, red, blue, green and magenta) are used in Fig. 8 to identify different phase and line integrated densities as following:

- bin 1: $\leq 0.6\pi$; $N_e \leq 3.9 \times 10^{18}$ cm$^{-3}$ (white),
- bin 2: $0.6\pi$ - $0.8\pi$; 3.9 - 5.7 $\times 10^{18}$ cm$^{-3}$ (red),
- bin 3: $0.8\pi$ - $1.1\pi$; 5.7 - 7.5 $\times 10^{18}$ cm$^{-3}$ (blue),
- bin 4: $1.1\pi$ - $1.3\pi$; 7.5 - 9.3 $\times 10^{18}$ cm$^{-3}$ (green),
- bin 5: $> 1.3\pi$ ; $> 9.3$  $10^{19}$ cm$^{-3}$ (magenta).

The variations of $<N_e>$ with the distance along the shock tube are reported in the right panel of Fig. 8 at 10 ns (in red), 20 ns (in blue), 30 ns (in green) and 40 ns (in magenta).

The interaction between the two precursors is clearly visible at 0.1 bar (Fig. 8(a)): at 10 ns, the interaction of the counter-propagating shocks has either not yet started or is below the sensitivity of the diagnostic. The interaction occurs at later times, with a typical signature which is as follows: the slope of $<N_e>$ is decreasing from the left (MAIN precursor), passes through a minimum and increases at the right (AUX). The minimum itself increases with time up to 7 $\times 10^{18}$ cm$^{-3}$ at 40 ns.

At 0.2 bar, the two records (with MAIN only and with the two shock waves) indicate a precursor for MAIN, with a slope which decreases from the left to the right in Fig. 8(b). Up to 40 ns, the precursors for the MAIN shock are very similar with and without (Fig. 8(c)) the presence of AUX shock wave and there is no obvious indication about a precursor for AUX in the case of two counter-propagating shock waves (Fig. 8(b)). At this pressure and compared with the previous case at 0.1 bar, the absence of precursor for AUX may be attributed to: i) a lower shock speed (18 km/s) combined with a larger pressure (hence the precursor extension increases with speed and decreases with pressure), ii) a too small longitudinal extension of the eventual precursor (see Fig. 8(b)) compared with the resolution of 20 microns (2 pixels). Our 1D numerical simulations with Xenon opacity multiplier $\times 20$ (not presented here), indicate a small precursor for AUX shock. At 15 ns, its extension is 50 $\mu$m (900 $\mu$m for MAIN shock) with a typical electron density $\sim 3.5 \times 10^{19}$ cm$^{-3}$ ($2.3 \times 10^{19}$ cm$^{-3}$ for MAIN shock), which does not agree with the record. At 42 ns the precursor of MAIN reaches the AUX shock front and the profile is similar to the profile at 20 ns shown in Fig. 3a at 0.1 bar, with a plateau of almost constant electron density between the two fronts. This might be compatible with small bending of the 4$^{th}$ fringe (from the right) between 45 and 50 ns. As 1D simulations are known to overestimate the precursor electron density, 2D simulations are necessary for a more precise interpretation of the experimental result.

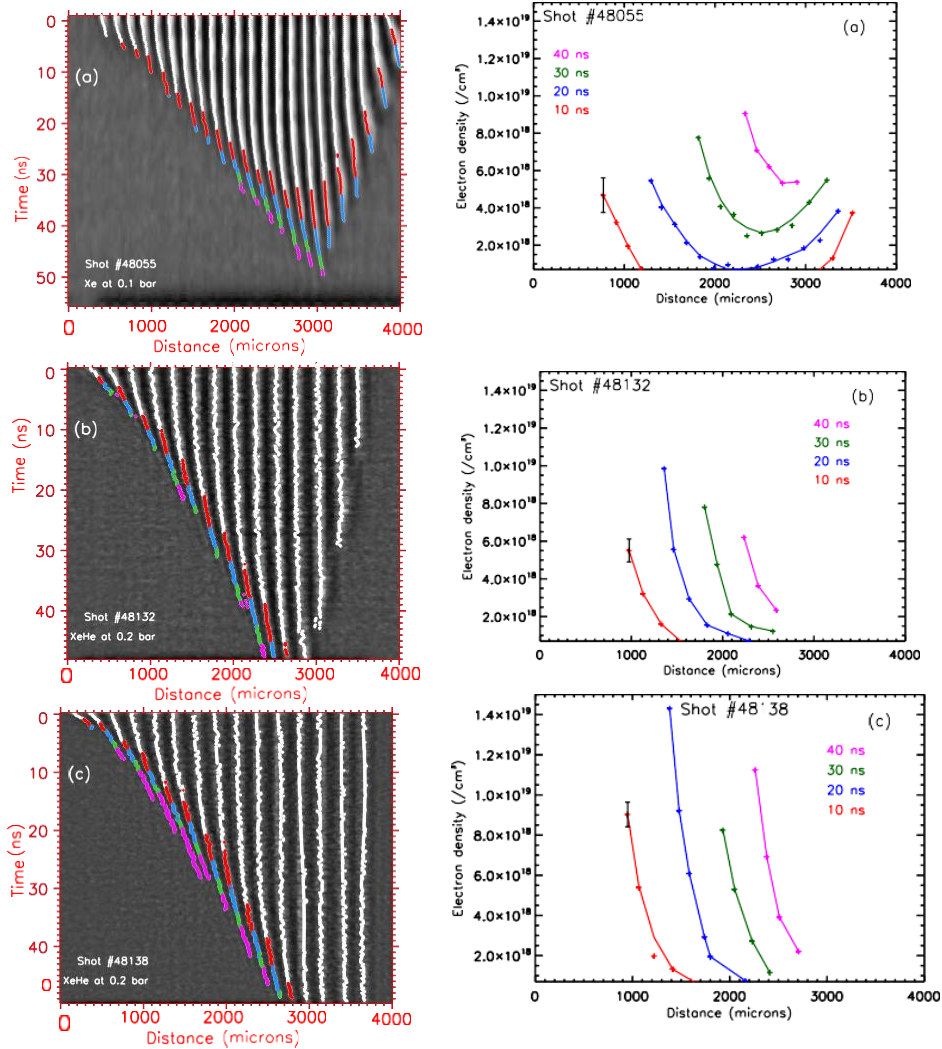

Figure 8: Left panel: interferometric records #48055 in Xe at 0.1 bar (a), #48132 in Xe+Heat 0.2 bar (b) and #48138 in Xe+He at 0.2 bar (c). Right panel: electron density at 10, 20, 30 and 40 ns versus distance for these records. The positions of maxima have been identified on the records in the left panel. The time t = 0, corresponds to the time of laser arrival on the target and the position x = 0 corresponds to the interface between the piston (Au layer) and the gas. Its determination is precise within 100 microns. The distances between two unperturbed fringes for records #48055, #48132 and #48138 are 159, 244 and 244 μm respectively. The $< N_e >$ uncertainty (2 pixels) is indicated by the error bar in the right panels. It corresponds respectively to $9 \times 10^{17}$, $6 \times 10^{17}$ and $6 \times 10^{17}$ cm$^{-3}$ for the figures (a), (b) and (c).

*4.3. XUV Spectroscopy*

Spectroscopy is an adequate tool to monitor the temperature of the plasma. It is thus appropriate for identifying the shocks collision by comparing the spectra for single and counter -propagating shock waves. Unfortunately, only a few records were obtained during the experiment preventing to catch the collision with this diagnostic. Among the shots recorded, the XUV spectrum of the shot #48143 is presented here is details. This shot was performed for [Xe (90%)+ He (10%)] mixture at 0.6 bar with laser energies of 123 J for MAIN and 63 J for AUX. The interferometric record of this shot is shown in Fig. 9a. The MAIN shock speed is estimated to be $39 \pm 4$ km/s. The estimation of the AUX speed ($18 \pm 5$ km/s) is imprecise, due to the presence of glue on the right section of the record. In this interferometric record, we have not been able to retrieve the collision time. However, if we extrapolate it for the speeds $\sim 39$ km/s (MAIN) and $\sim 18$ km/s (AUX), this collision should happen between 60 to 65 ns with an increase of the temperature up to 30 eV.

The raw spectrum recorded for the wavelength range of 15-35 nm with the L edge of Aluminum at 17 nm (34 nm in second order). The net spectrum, corrected from the transmission [29] of the 100 nm thick $Si_3N_4$ window (3 mm × 0.4 mm) is presented in Fig. 9b. A remarkable feature is a strong absorption dip between 19 and 22 nm. This absorption probably comes from the cold layer (thickness 300 $\mu$m) between the shock heated plasma and the $Si_3N_4$ window.

Few lines of Xe VII-VIII are identified through NIST database [1] as also Oxygen IV and V lines. Lyman lines of He II (from 1-2 to 1-7) are also present in the spectrum. These informations will be useful for estimation of the electron temperature (section 4.4).

*4.4. 1D simulations based on experimental results*

We will now compare the experimental shock characteristics with the results of HELIOS 1D simulations using the PROPACEOS equation of state and opacity (limited to 1 group). This opacity has been multiplied by 20 for Xenon. As our interest is to understand the shock structure in Xenon and not the laser matter interaction on the piston, we adjusted the fluences in order to reproduce the experimental speeds.

To analyse the results from the shot #48055 (Fig. 8(a)), we thus set the fluences to 32000 and 7500 J/cm$^2$. This allows producing the experimental shock speeds 54 and 23 km/s in Xenon at 0.1 bar for the MAIN and AUX beams, respectively. The two shocks appear in Xenon at 2 and 3 ns respectively for MAIN and AUX. The merging of the two precursors starts at ∼ 15 ns and the shock collision time occurs at 47 ns. In Fig. 10, we present the electron density profiles from the simulation (dotted lines) and the experiment (solid lines) at 10, 20, 30 and 40 ns.

At 10 ns, the two simulated precursor extensions are 0.165 and 0.022 cm for MAIN and AUX respectively. The electron density is larger by a factor of 4 than in the experiment. The shapes of the precursors are also very different. However, the simulation supposes that the plasma is uniform in the transverse direction (i.e. perpendicular to the shock propagation) whereas, in the experiment, N$_e$ decreases from the center axis to the walls, due to radiation cooling.

---

[1][http://physics.nist.gov/PhysRefData/ASD/lines_form.html]

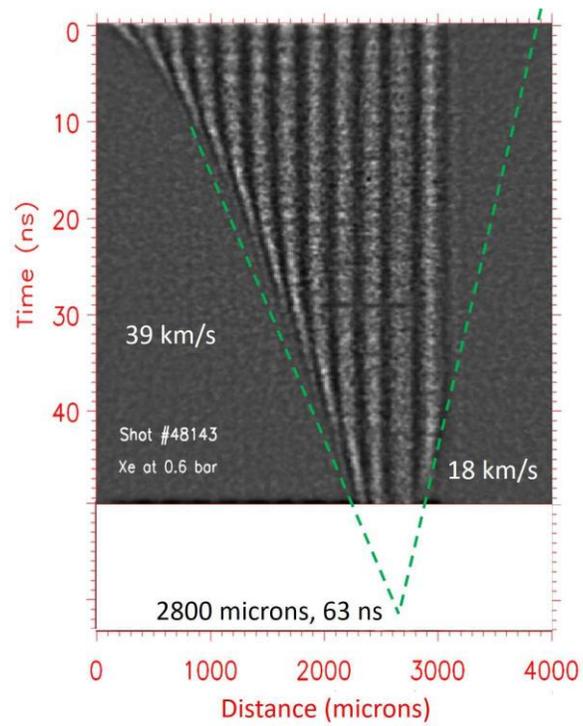

(a)

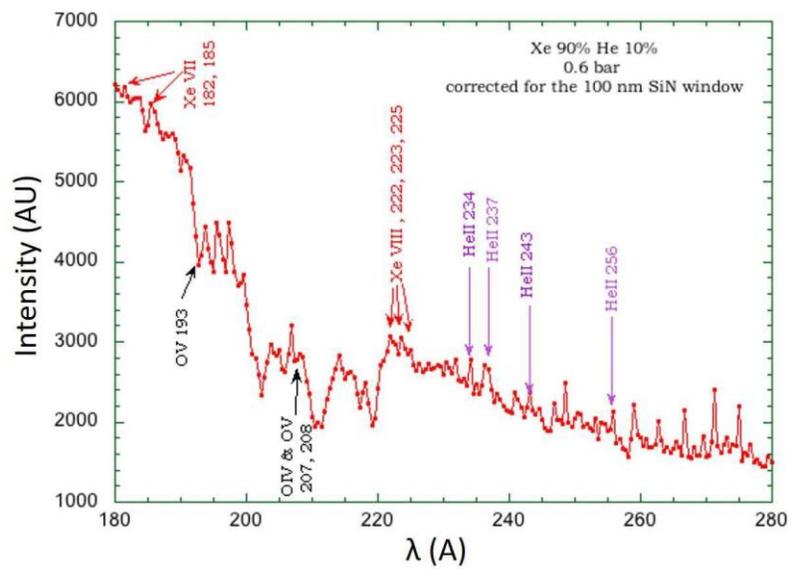

(b)

Figure 9: Visible interferometric image (a) and XUV spectrum (b) recorded for the shot#48143.

As a consequence, the values of the experimental line averaged $N_e$ should be smaller than the numerical ones and the corresponding profiles should be smoother. It is also important to note that, for AUX shock, the average $<N_e>$ value underestimates the local value at the centre of the tube by at least a factor of about 2 (as it is averaged over 0.6 mm instead of 0.3 mm). Moreover, our 1D simulation suffers from an inexact opacity and 2D effects are probably important especially for AUX. Thus we have here only a qualitative interpretation of the experimental results.

The interaction between the two HELIOS radiative precursors starts between 10 and 20 ns, like in the experiment. However, the shape, as well as absolute values of the simulated electron density curves, are not in agreement with the experimental results and the interaction is stronger in the simulation than in the experiment.

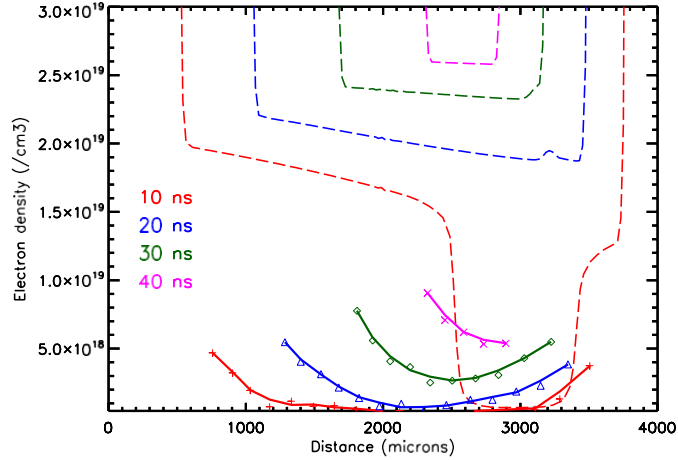

Figure 10: Recorded electron density (shot #48055) together with the HELIOS results at different times.

In order to interpret the spectroscopic data presented in section 4.3, we performed another 1D simulation in Xenon at 0.6 bar, and adapted the fluences to generate two counter-propagating shocks with the speeds 36 and 18 km/s, close to the experiment. The time evolutions of the electron density, mean

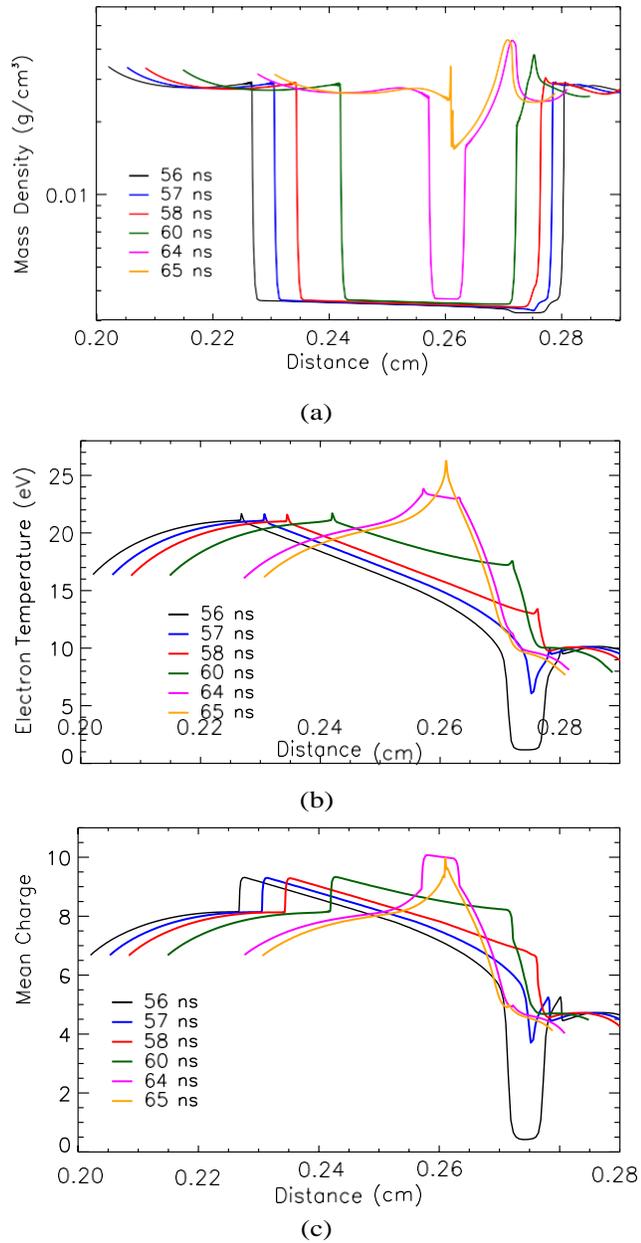

Figure 11: Time evolution of the mass density (a), electron temperature (b) and mean charge (c) at 56, 57, 58, 60, 64 and 65 ns within the shock tube derived from the HELIOS simulations(with Xenon opacity multiplier = 20), for two counter streaming shocks of ∼ 39 and 18 km/s.

charge and electron temperature at 56, 57, 58, 60, 64 and 65 ns are presented in Fig. 11.

The two shocks appear in Xenon at 2 and 3 ns respectively for MAIN and AUX. Concerning AUX, the combination of a small speed and a relatively high pressure does not allow to develop a radiative precursor, in agreement with the experimental results (Fig. 9a), whereas the MAIN shock has a precursor and its length is increasing with time. The post-shock temperature of the MAIN is ∼ 21 eV and the ion charge ∼ 9. At 57 ns, the precursor of MAIN reaches the AUX shock front. This time is out of our record (see Fig. 9a) which means that the interaction effect is either absent or occurs at later times. The structure of the AUX post shock is modified by the interaction with the MAIN precursor (Fig. 11b). The shock collision occurs at 65 ns (Fig. 11a) resulting in the development of two reserve shock waves. At the collision time, the electron density, mass density, electron temperature and ion charge reach respectively ∼ $1.4 \times 10^{21}$ cm$^{-3}$, 0.034 g/cm$^{-3}$, 26 eV and 10.

In order to interpret the spectroscopic results, qualitative preliminary computations of the XUV spectra emerging from a 600 $\mu$m thick plasma with two representative values of the mass density, $\rho = 3.2 \times 10^{-2}$ and $3.3 \times 10^{-3}$ g/cm$^3$ have been performed at different temperatures, using the methods described in [28]. They show that the lines of HeII can only be observed at a temperature of ∼ 15 eV and for the lowest density, i.e. in the radiative precursor. On another side, the presence of lines of Xe VII-VIII in the record is compatible with our 1D simulations.

5. Discussion and conclusions

This paper reports the first experimental study of the interaction between two radiative shock waves propagating at two different speeds in Xenon. This interaction has been analysed by optical interferometry, XUV spectroscopy and interpreted by 1D simulations. At 0.1 bar and at 54 and 23 km/s, the interaction is clearly characterized in the experiment by the enhancement of the ionisation wave followed by the merging of the two radiative precursors at 20 ns. The collision time is recorded at 47 ns. Such behavior is reproduced by the simulation. However, the interaction effect is larger in the simulation than in the experiment.

We have investigated this interaction at larger pressure, 0.2 bar, with the following speeds ∼ 41 km/s for MAIN and ∼ 18 km/s for AUX shock waves. We do not record any experimental signature of the radiative precursor for AUX, and we do not catch experimentally the collision time. The recorded precursor of MAIN is not influenced by the AUX wave up to 40 ns (Fig. 8(b) and (c)) which is the limit of the record. On its side, the 1D simulation predicts a tiny precursor for AUX. At 42 ns the precursor of MAIN reaches the AUX shock front, resulting in a flat profile of $N_e$.

The results of the transverse interferometry at 0.2 bar, with speeds of ∼ 40 and 20 km/s indicate that the MAIN precursor has a lateral extension of ∼ 600 $\mu$m whereas it is 300 $\mu$m for AUX. The precursor of MAIN is almost flat with a probable small bending at the edges of the tube, whereas the AUX precursor is more curved. This means that the 2D effects are more important for AUX than for MAIN.

Our simulations give a qualitative description of the shocks interaction when the laser fluence is adjusted to give the correct shock velocities. However, it is now well known that 2D simulations (together with state of the art opacities) fit better with experiments [15, 14, 8]. For the same laser energy, 2D simulations lead to a diminution of the shock speed compared to 1D as also to a diminution of the electron density. For instance, in the case of a shock wave launched by a laser beam at 1315 nm in Xenon at 0.3 bar at PALS and with a laser fluence of 85000 J/cm$^2$, ARWEN 2D simulations give a shock speed of 44 km/s in agreement with the experimental one [30]. 1D simulation would require in this case a fluence of 30000 J/cm$^2$ to achieve the same.

The space and time integrated XUV records at 0.6 bar, for respective speeds which are equal to ∼ 39 and 18 km/s, indicate that the temperature of the shock has reached values up to 15 eV and that the Xenon mean ion charge has also reached values of 6 - 7 whereas 1D simulations predict electron temperature 10 - 30 eV and ion charge 5 - 10 (Fig. 11c). A more detailed study, based on 2D simulation and radiative transfer post-processing will be necessary refine in the analysis.

## 6. Appendix

The quality of the LTE Planck and Rosseland opacities is crucial for a precise numerical modeling of the radiative shocks, especially those presenting a developed precursor. The calculation of these functions requires the determination of the monochromatic opacity in the relevant part of the two weighting functions entering in the average, which peak at 2.8 and 3.8 kT respectively. For our experimental radiative precursors, which have a constant mass density $\rho = 5.1 \times 10^{-4}$ g/cm$^3$, the two maximum expected temperatures are around

15 and 7 eV, which means that the opacities have to be accurate between few eV to 15 eV. The continuum lowering is introduced within the ion sphere model [31] for temperatures lower than 6 eV and a StewartPyatt [32] model for higher temperatures. We found a good agreement with ATOMIC [33, 34] results especially at the lowest temperatures, where the modeling remains delicate. More details may be found in [28].

The mean ion charges, computed by the two methods, are in excellent agreement, except at very low temperature (Fig.12). However, the Planck and Rosseland opacities differ noticeably as can be seen from the Fig. 13a and 13b. Near 10 eV, which corresponds to the precursor plateau, our opacity is greater than the PROPACEOS one by a factor of 7 for the Rosseland and 25 for the Planck.

In order to keep this difference into account, the only possibility was to use a multiplicative factor for the two opacities, which was then taken 20. We are then aware that the present HELIOS results have to be used only for a qualitative interpretation.

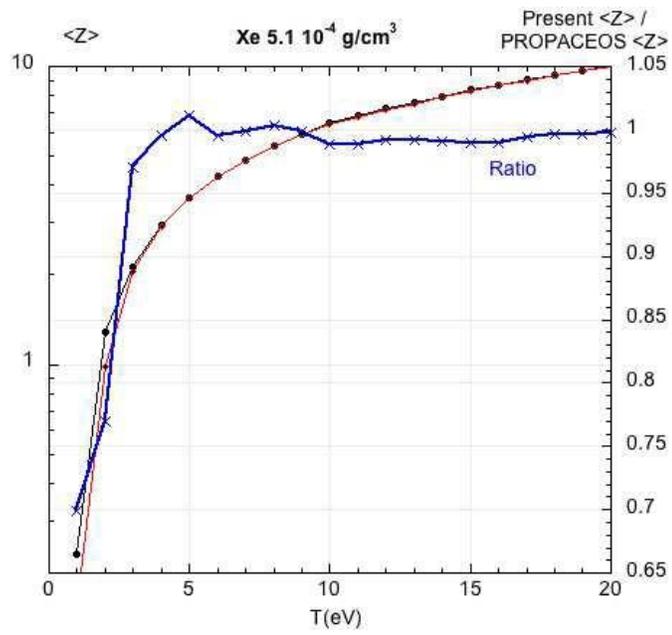

Figure 12: mean ion charge for Xenon at $5.1 \times 10^{-4}$ g/cm$^3$, for PROPACEOS and our model. The ratio $<z>_{ourmodel}/<z>_{PROPACEOS}$ is also plotted in blue.

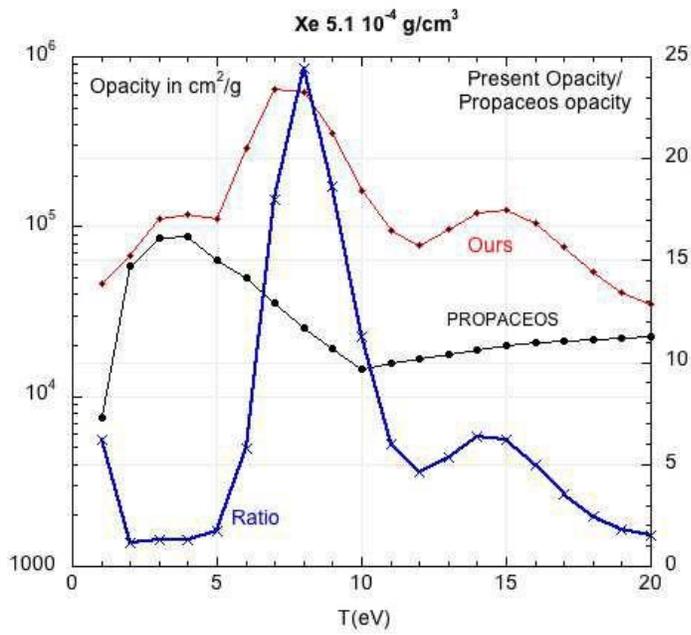

(a)

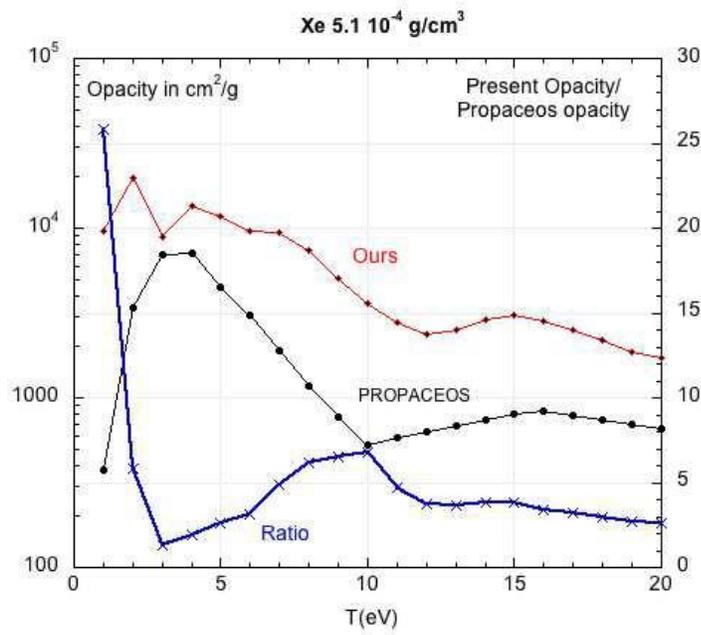

(b)

Figure 13: Planck (a) and Rosseland (b) opacities in the conditions of Fig.12. In red: our model; in black: PROPACEOS. The ratio (in blue) between the two is also plotted in linear scale.

7. Acknowledgements

We are grateful to J. Golasowski, J. Hrebicek, T. Medrik for their help in the diagnostics, to S. Croce, T. Mesle, F. Reix, Y. Younes, and P. Jagourel, from Pole Instrumental de lObservatoire de Paris for the targets manufacturing, to Chris Spindloe (Scitech) for the gilt parylene foils, to Roland Smith (Imperial College) for the Dove prism, and to J. Skala, E. Krousky and M. Pfeifer from PALS for their assistance. Finally, we acknowledge the support of the Scientic Council of Observatoire de Paris, the French CNRS/INSU Programme PNPS, the PICS 6838 of CNRS, the Labex PLAS@PAR (ANR-11-IDEX-0004-02), the UFR de Physique of UPMC as also of EXTREME LIGHT INFRASTRUCTURE project no. CZ.1.05/1.1.00/02.0061 and EC OP projects no. CZ.1.07/2.3.00/30.0057 and CZ.1.07/2.3.00/20.0279, which were co- financed 285 by the European Social Funds and the Ministry of Education, Youth and Sports of the Czech Republic - projects LM2010014 and LM2015083 (PALS RI), and the LASERLABEUROPE (grant agreement no. 284464, ECs Seventh Framework Programme).

References

[1] R. P. Drake, L. Davison, Y. Horie, High-Energy-Density Physics: Fundamentals, Inertial Fusion, and Experimental Astrophysics, Springer, 2006.

[2] D. Mihalas, B. Weibel-Mihalas, Foundations of radiation hydrodynamics, Dover Publications, 1984.

[3] C. Michaut, L. Boireau, T. Vinci, S. Bouquet, M. Koenig, A. Benuzzi-Mounaix, N. Ozaki, C. Clique, S. Atzeni, Experimental and numerical studies of radiative shocks, Journal de Physique IV 133 (2006) 1013–1017. doi:10.1051/jp4:2006133205.

[4] S. Bouquet, C. Stehlé, M. Koenig, J.-P. Chièze, A. Benuzzi-Mounaix, D. Batani, S. Leygnac, X. Fleury, H. Merdji, C. Michaut, F. Thais,

N. Grandjouan, T. Hall, E. Henry, V. Malka, J.-P. J. Lafon, Observation of Laser Driven Supercritical Radiative Shock Precursors, Physical Review Letters 92 (22) (2004) 225001. doi:10.1103/PhysRevLett.92.225001.

[5] U. Chaulagain, C. Stehlé, J. Larour, M. Kozlová, F. Suzuki-Vidal, P. Barroso, M. Cotelo, P. Velarde, R. Rodriguez, J. M. Gil, A. Ciardi, O. Acef, J. Nejdl, L. de Sá, R. L. Singh, L. Ibgui, N. Champion, Structure of a laser-driven radiative shock, High Energy Density Physics 17A (2015) 106–113. doi:10.1016/j.hedp.2015.01.003.

[6] M. González, C. Stehlé, M. Busquet, B. Rus, F. Thais, O. Acef, P. Barroso, A. Bar-Shalom, D. Bauduin, M. Kozlova, T. Lery, A. Madouri, T. Mocek, J. Polan, Astrophysical radiative shocks: From modeling to laboratory experiments, Laser and Particle Beams 24 (2006) 535540. doi:10.1017/s026303460606071x.

[7] A. B. Reighard, R. P. Drake, K. K. Dannenberg, D. J. Kremer, M. Grosskopf, E. C. Harding, D. R. Leibrandt, S. G. Glendinning, T. S. Perry, B. A. Remington, J. Greenough, J. Knauer, T. Boehly, S. Bouquet, L. Boireau, M. Koenig, T. Vinci, Observation of collapsing radiative shocks in laboratory experiments, Physics of Plasmas 13 (8) (2006) 082901. doi:10.1063/1.2222294.

[8] C. Stehlé, M. González, M. Kozlova, B. Rus, T. Mocek, O. Acef, J. P. Colombier, T. Lanz, N. Champion, K. Jakubczak, J. Polan, P. Barroso, D. Bauduin, E. Audit, J. Dostal, M. Stupka, Experimental study of radiative shocks at PALS facility, Laser and Particle Beams 28 (2010) 253–261. arXiv:1003.2739, doi:10.1017/S0263034610000121.

[9] F. W. Doss, H. F. Robey, R. P. Drake, C. C. Kuranz, Wall shocks in high-energy-density shock tube experiments, Physics of Plasmas 16 (11) (2009) 112705. doi:10.1063/1.3259354.

[10] R. P. Drake, F. W. Doss, R. G. McClarren, M. L. Adams, N. Amato, D. Bingham, C. C. Chou, C. DiStefano, K. Fidkowski, B. Fryxell, T. I.

Gombosi, M. J. Grosskopf, J. P. Holloway, B. van der Holst, C. M. Huntington, S. Karni, C. M. Krauland, C. C. Kuranz, E. Larsen, B. van Leer, B. Mallick, D. Marion, W. Martin, J. E. Morel, E. S. Myra, V. Nair, K. G. Powell, L. Rauchwerger, P. Roe, E. Rutter, I. V. Sokolov, Q. Stout, B. R. Torralva, G. Toth, K. Thornton, A. J. Visco, Radiative effects in radia- tive shocks in shock tubes, High Energy Density Physics 7 (2011) 130–140. doi:10.1016/j.hedp.2011.03.005.

[11] A. Dizière, C. Michaut, M. Koenig, C. D. Gregory, A. Ravasio, Y. Sakawa, Y. Kuramitsu, T. Morita, T. Ide, H. Tanji, H. Takabe, P. Barroso, J.-M. Boudenne, Highly radiative shock experiments driven by GEKKO XII, *Astrophys. Space Sci.* 336 (2011) 213–218. doi:10.1007/s10509-011-0653-6.

[12] C. Stehlé, M. Kozlová, J. Larour, J. Nejdl, N. Champion, P. Barroso, F. Suzuki-Vidal, O. Acef, P.-A. Delattre, J. Dostál, M. Krus, J.-P. Chièze, New probing techniques of radiative shocks, Optics Communications 285 (2012) 64–69. doi:10.1016/j.optcom.2011.09.008.

[13] J. E. Bailey, G. A. Chandler, S. A. Slutz, I. Golovkin, P. W. Lake, J. J. MacFarlane, R. C. Mancini, T. J. Burris-Mog, G. Cooper, R. J. Leeper, T. A. Mehlhorn, T. C. Moore, T. J. Nash, D. S. Nielsen, C. L. Ruiz, D. G. Schroen, W. A. Varnum, Hot dense capsule-implosion cores produced by z-pinch dynamic hohlraum radiation, Phys. Rev. Lett. 92 (2004) 085002. doi:10.1103/PhysRevLett.92.085002.

[14] S. Leygnac, L. Boireau, C. Michaut, T. Lanz, C. Stehlé, C. Clique, S. Bouquet, Modeling multidimensional effects in the propaga- tion of radiative shocks, Physics of Plasmas 13 (11) (2006) 113301. doi:10.1063/1.2366544.

[15] M. González, E. Audit, C. Stehlé, 2D numerical study of the radiation influence on shock structure relevant to laboratory astrophysics , *Astron. Astrophys.* 497 (2009) 27–34. doi:10.1051/0004-6361/20079136.

[16] F. W. Doss, H. F. Robey, R. P. Drake, C. C. Kuranz, Wall shocks in high-energy-density shock tube experiments, Physics of Plasmas 16 (11) (2009) 112705. doi:10.1063/1.3259354.

[17] C. F. McKee, X-Ray Emission from an Inward-Propagating Shockin Young Supernova Remnants, *Astrophys. J.*188 (1974) 335–340. doi:10.1086/152721.

[18] R. A. Chevalier, Self-similar solutions for the interaction of stellar ejecta with an external medium, *Astrophys. J.*258 (1982) 790–797. Doi:10.1086/160126.

[19] S. Orlando, R. Bonito, C. Argiroffi, F. Reale, G. Peres, M. Miceli, T. Matsakos, C. Stehlé, L. Ibgui, L. de Sa, J. P. Chièze, T. Lanz, Radiative accretion shocks along nonuniform stellar magnetic fields in classical T Tauri stars, *Astron. Astrophys.*559 (2013) A127. doi:10.1051/0004-6361/201322076.

[20] P. Hartigan, The visibility of the Mach disk and the bow shock of a stellar jet, *Astrophys. J.*339 (1989) 987–999. doi:10.1086/167353.

[21] A. C. Raga, G. Mellema, S. J. Arthur, L. Binette, P. Ferruit, W. Steffen, 3D Transfer of the Diffuse Ionizing Radiation in ISM Flows and the Preionization of a Herbig-Haro Working Surface, *Revista Mexicana de Astronomia y Astrofisica*35 (1999) 123–133.

[22] R. M. Williams, Y.-H. Chu, J. R. Dickel, R. Beyer, R. Petre, R. C. Smith, D. K. Milne, Supernova Remnants in the Magellanic Clouds. I. The Colliding Remnants DEM L316, *Astrophys. J.*480 (1997) 618–632.

[23] R. M. Williams, Y.-H. Chu, J. R. Dickel, R. A. Gruendl, F. D. Seward, M. A. Guerrero, G. Hobbs, Supernova Remnants in the Magellanic Clouds. V. The Complex Interior Structure of the N206 Supernova Remnant, *Astrophys. J.*628 (2005) 704–720. doi:10.1086/431349.

[24] J. C. Toledo-Roy, P. F. Velázquez, F. de Colle, R. F. González, E. M. Reynoso, S. E. Kurtz, J. Reyes-Iturbide, Numerical model for the SNR

DEM L316: simulated X-ray emission, *Mon. Not. Roy. Astron. Soc.* 395 (2009) 351–357. doi:10.1111/j.1365-2966.2009.14517.x.

[25] P. Velarde, D. García-Senz, E. Bravo, F. Ogando, A. Relaño, C. García, E. Oliva, Interaction of supernova remnants: From the circumstellar medium to the terrestrial laboratory, Physics of Plasmas 13 (9) (2006) 0929011. doi:10.1063/1.2338281.

[26] K. Jungwirth, A. Cejnarova, L. Juha, B. Kralikova, J. Krasa, E. Krousky, P. Krupickova, L. Laska, K. Masek, T. Mocek, M. Pfeifer, A. Präg, O. Renner, K. Rohlena, B. Rus, J. Skala, P. Straka, J. Ullschmied, The Prague Asterix Laser System, Physics of Plasmas 8 (2001) 2495–2501. doi:10.1063/1.1350569.

[27] J. J. MacFarlane, I. E. Golovkin, P. R. Woodruff, HELIOS-CR A 1-D radiation-magnetohydrodynamics code with inline atomic kinetics modeling, *Journal of Quantitative Spectroscopy & Radiative Transfer* 99 (2006) 381-397. doi:10.1016/j.jqsrt.2005.05.031.

[28] R. Rodríguez, G. Espinosa, J. M. Gil, C. Stehlé, F. Suzuki-Vidal, J. G. Rubiano, P. Martel, E. M´ınguez, Microscopic properties of xenon plasmas for density and temperature regimes of laboratory astrophysics experiments on radiative shocks, *Phys. Rev. E* 91 (5) (2015) 053106. doi:10.1103/PhysRevE.91.053106.

[29] B. L. Henke, E. M. Gullikson, J. C. Davis, X-Ray Interactions: Photoabsorption, Scattering, Transmission, and Reflection at E = 50-30,000 eV, Z = 1-92, Atomic Data and Nuclear Data Tables 54 (1993) 181–342. doi:10.1006/adnd.1993.1013.

[30] M. Cotelo, P. Velarde, A. G. de la Varga, D. Portillo, C. Stehlé, U. Chaulagain, M. Kozlova, J. Larour, F. Suzuki-Vidal, Simulation of radiative shock waves in Xe of last PALS experiments, High Energy Density Physics 17A (2015) 68–73. doi:10.1016/j.hedp.2014.12.002.

[31] S. Ichimaru, Strongly coupled plasmas: high-density classical plasmas and degenerate electron liquids, Reviews of Modern Physics 54 (1982) 1017–1059. `doi:10.1103/RevModPhys.54.1017`.

[32] J. C. Stewart, K. D. Pyatt, Jr., Lowering of Ionization Potentials in Plasmas, *Astrophys. J.* 144 (1966) 1203. `doi:10.1086/148714`.

[33] C. J. Fontes, J. Colgan, H. L. Zhang, J. Abdallah, Large-scale kinetics modeling of non-LTE plasmas, *Journal of Quantitative Spectroscopy & Radiative Transfer* 99 (2006) 175–185. `doi:10.1016/j.jqsrt.2005.05.014`.

[34] M. F. Gu, The flexible atomic code, Canadian Journal of Physics 86 (2008) 675–689. `doi:10.1139/P07-197`.